\documentclass[twocolumn]{openjournal}
\usepackage{soul}
\usepackage{natbib}
\usepackage{hyperref}
\usepackage{amsmath}
\usepackage{xspace}
\usepackage{lineno}
\usepackage{graphicx}
\usepackage{enumitem}
\usepackage{amssymb}
\usepackage{xifthen}
\usepackage[normalem]{ulem}
\usepackage{orcidlink}
\usepackage{longtable}
\usepackage{booktabs} 
\usepackage{xcolor}

\definecolor{linkblue}{RGB}{0, 70, 180}
\hypersetup{
colorlinks = true,
linkcolor  = linkblue,
citecolor  = linkblue,
urlcolor   = linkblue,
}

\newcommand{\software}[1]{\par\noindent\textit{Software:} #1}
\newcommand{\facilities}[1]{\par\noindent\textit{Facilities:} #1}
\newcommand{\okina}{\textquoteleft}

\shorttitle{MDF of Segue~1}
\shortauthors{Bissonette et al.}

\begin{document}

\title{The Metallicity Distribution of the Ultra-Faint Dwarf Galaxy Segue~1}

\author{Daisy~Bissonette\,\orcidlink{0009-0006-5977-618X}$^{1,2}$}
\author{Alexander~P.~Ji\,\orcidlink{0000-0002-4863-8842}$^{1,2,3}$}
\author{Joshua D. Simon\,\orcidlink{}$^{4}$}
\author{Joss Bland-Hawthorn\,\orcidlink{0000-0001-7516-4016}$^{5}$} 
\author{Anirudh Chiti\,\orcidlink{0000-0002-7155-679X}$^{6,7}$}
\author{Marla Geha\,\orcidlink{}$^{8}$}
\author{Ting S. Li\,\orcidlink{0000-0002-9110-6163}$^{9,10}$}
\author{Anna Frebel\,\orcidlink{}$^{11}$} 
\author{Alice~M.~Luna\,\orcidlink{0009-0009-9570-0715}$^{1,2}$} 

\affiliation{$^{1}$\textit{Department of Astronomy \& Astrophysics, University of Chicago, 5640 S Ellis Avenue, Chicago, IL 60637, USA}}
\affiliation{$^{2}$\textit{Kavli Institute for Cosmological Physics, University of Chicago, Chicago, IL 60637, USA}}
\affiliation{$^{3}$\textit{NSF-Simons AI Institute for the Sky (SkAI), 172 E. Chestnut St., Chicago, IL 60611, USA}}
\affiliation{$^{4}$\textit{Observatories of the Carnegie Institution for Science, 813 Santa Barbara St., Pasadena, CA 91101, USA}}
\affiliation{$^{5}$\textit{Sydney Institute for Astronomy, School of Physics, A28, The University of Sydney, NSW 2006, Australia}}
\affiliation{$^{6}$\textit{Kavli Institute for Particle Astrophysics \& Cosmology, Stanford University, Stanford, CA 94305, USA}}
\affiliation{$^{7}$\textit{Brinson Prize Fellow}}
\affiliation{$^{8}$\textit{Department of Astronomy, Yale University, New Haven, CT 06520, USA}}
\affiliation{$^{9}$\textit{Department of Astronomy and Astrophysics, University of Toronto, 50 St. George Street, Toronto, ON, M5S 3H4, Canada}}
\affiliation{$^{10}$\textit{Dunlap Institute for Astronomy \& Astrophysics, 50 St. George Street, Toronto, ON, M5S 3H4, Canada}}
\affiliation{$^{11}$\textit{Department of Physics and Kavli Institute for Astrophysics and Space Research, Massachusetts Institute of Technology, 77 Massachusetts Avenue, Cambridge, MA 02139, USA}}

\begin{abstract}

Ultra-faint dwarf galaxies (UFDs; stellar masses $M_* < 10^5\,M_\odot$) offer unique insights into early chemical evolution in low-mass systems. However, interpreting their metallicity distribution functions (MDFs) has been challenging due to limited spectroscopic samples, especially beyond the red giant branch. We present metallicities from the Ca~II~K absorption feature, measured from low-resolution ($R\sim1000$) Keck/LRIS spectroscopy of 40 stars in the UFD Segue~1 ($M_* \approx 500\,M_\odot$), including both red giant branch and main-sequence turnoff stars, resulting in a metallicity sample more than six times larger than previously published data for Segue~1. The resulting MDF has an average [Fe/H] $= -2.52 \pm 0.10$ dex and a dispersion of $\sigma = 0.59 \pm 0.06$ dex, and is best described by a single Gaussian, with no evidence for multimodal substructure. This is consistent with a continuous, short-duration ($\lesssim$1~Gyr) episode of star formation and chemical enrichment prior to reionization. The nonzero metallicity spread reaffirms its classification as a galaxy. Segue~1 highlights the rich chemical enrichment histories present even in the least massive galaxies, and underscores the importance of deep spectroscopic follow-up to fully characterize these ancient stellar systems.

\end{abstract}


\section{Introduction}\label{sec:intro}

Ultra-faint dwarf galaxies (UFDs), defined as the faintest Milky Way satellites confirmed to be galaxies ($L_V < 10^{5}$ L\textsubscript{$\odot$}), are among the most dark-matter-dominated systems known, with extremely high mass-to-light ratios and the lowest average metallicities of any stellar population \citep{simon_kinematics_2007, belokurov_cats_2007, simon_faintest_2019, geha_deimos2_2026}. Thought to be relics of the early universe, UFDs offer a unique window into the first galaxies and the role of dark matter on small scales. Their low stellar masses suggest only a few brief bursts of star formation, with most stars forming within the first $\sim$~1~Gyr of cosmic history before reionization (z~$\sim$~6) quenched star formation \citep{brown_quenching_2014, rodriguezwimberly_suppression_2019}. However, recent work suggests that some UFDs may still have formed stars after reionization \citep{savino_sfh_2023,mcquinn_sfh_2024,durbin_hst_2025,luna_bimodal_2025}.

\begin{figure}[htbp]
      \centering
      \includegraphics[width=\columnwidth]{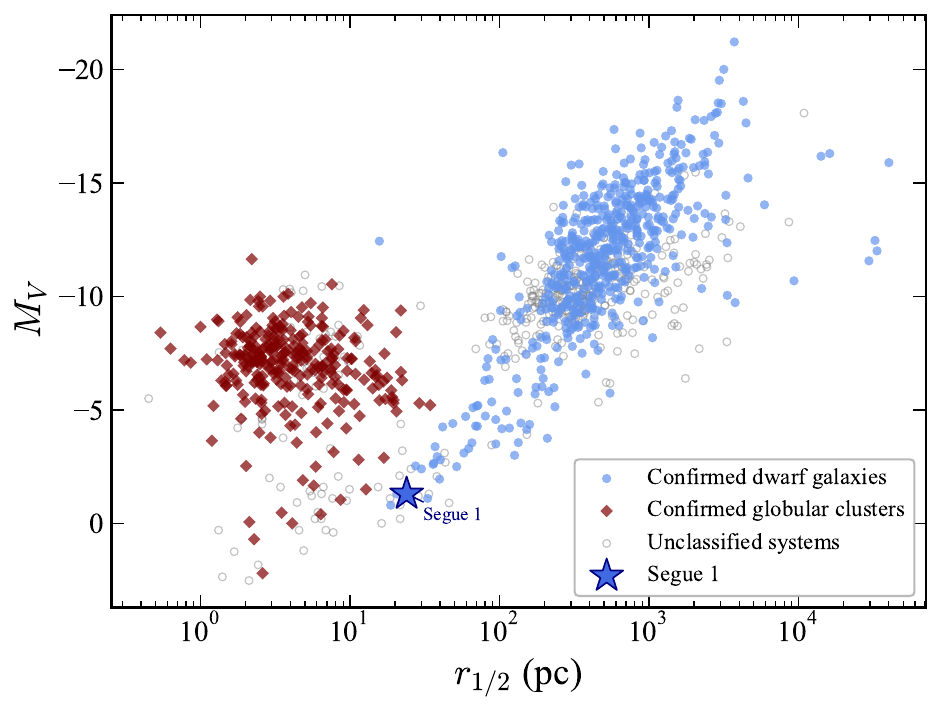}
      \caption{Absolute $V$-band magnitude ($M_V$) versus projected half-light radius ($r_{1/2}$) for dwarf galaxies (blue circles), globular clusters (red diamonds), and unclassified systems (open gray circles) in the Local Volume Database \citep[LVD;][v1.1.0]{pace_local_2025}, with $M_V$ computed from the tabulated apparent $V$-band magnitudes and distance moduli. Segue~1 is highlighted as a blue star. The clear separation between the compact globular cluster sequence and the more extended dwarf galaxy population illustrates the structural distinction between these two classes of stellar system; Segue~1 occupies the ultra-faint dwarf regime, with properties consistent with a dark matter-dominated satellite.}
      \label{fig:mv_rhalf}
\end{figure}

At these faint limits, distinguishing between galaxies and star clusters -- gravitationally bound stellar systems lacking significant dark matter or metallicity spread -- becomes challenging. This is a foundational issue in astrophysics, as it directly informs our understanding of how the smallest dark matter halos form, retain gas, and sustain chemical enrichment in the early universe \citep{bromm_first_2011}. Several of these systems, including Segue~1, Willman~1, Tucana~III, DELVE~1, Eridanus~III, and Ursa Major~III/UNIONS~1, occupy a previously underpopulated region of the size--luminosity plane, with half-light radii ($r_{1/2} \lesssim 30$~pc) and luminosities ($M_V \gtrsim -3$) that overlap with both faint globular clusters and ultra-faint dwarf galaxies (Figure~\ref{fig:mv_rhalf}). This overlap makes their initial classification ambiguous \citep{willman_new_2005, willman_new_2005-1, belokurov_cats_2007, geha_least-luminous_2009, drlica-wagner_eight_2015, simon_eridanus_2024, cerny_no_2025}.  

\citet{willman_galaxy_2012} proposed that an object should be classified as a galaxy if it shows dynamical evidence of dark matter and/or a chemical enrichment history implying retention of supernova ejecta within a deep potential well. Systems that fall below this threshold, lacking dark matter and showing no metallicity spread, are more consistent with star clusters. Additionally, recent work has reinforced that a physical size cut of $\sim$15~pc remains an effective empirical criterion for distinguishing star clusters from galaxies \citep{cerny_chemodynamical_2026, geha_deimos_2026}. Identifying which faint stellar systems qualify as galaxies is essential for tracing the threshold of galaxy formation and understanding how feedback, dark matter, and chemical evolution operate in the smallest galactic building blocks.

Segue~1 lies at the boundary between dwarf galaxies and globular clusters, as seen in Figure~\ref{fig:mv_rhalf}, making it a key system for applying the dynamical and chemical criteria proposed by \citet{willman_galaxy_2012}. Segue~1 is a UFD that was first discovered in the Sloan Digital Sky Survey (SDSS, \citealt{york_sloan_2000}) in 2007, though its classification as a UFD was initially disputed \citep{belokurov_cats_2007, niederste-ostholt_origin_2009}. Later, Segue~1 was determined to be a galaxy through kinematic and spectroscopic observations showing a velocity dispersion inconsistent with a purely stellar system \citep{geha_least-luminous_2009, simon_complete_2011}. With a stellar mass of only $\sim$500\(M_\odot\), a half-light radius of 24 pc, and a mass-to-light ratio of 2460~$M_{\odot}/L_{\odot}$, Segue~1 is the lowest-stellar-mass confirmed dwarf galaxy with a resolved velocity dispersion \citep{geha_least-luminous_2009, simon_complete_2011, frebel_segue_2014, munoz_megacam_2018}. Table~\ref{tab:segue1_properties} lists key properties of Segue~1, including its size, velocity dispersion, and inferred dark matter halo mass.

\begin{table}[h!]
\centering
\caption{Summary of Properties of Segue~1}
\label{tab:segue1_properties}
\renewcommand{\arraystretch}{1.3}
\begin{tabular}{lll}
    \toprule
    Row & Quantity & Value \\
    \midrule
    (1) & R.A. (J2000) (h m s) & 10:07:00.10 \\
    (2) & Decl. (J2000) ($^\circ$ ' '') & +16:04:32.2 \\
    (3) & Distance (kpc) & 23.0 $\pm$ 2 \\
    (4) & $M_V$ & $-$1.30 $\pm$ 0.73 \\
    (5) & $L_V (L_\odot)$ & 283 \\
    (6) & $\epsilon$ & 0.33$^{+0.10}_{-0.10}$ \\
    (7) & $\mu_{V,0}$ (mag arcsec$^{-2}$) & 28.06$^{+1.01}_{-0.98}$ \\
    (8) & $r_{\text{\textit{h,p}}}$ (pc) & 24.2$^{+2.8}_{-2.8}$ \\
    (9) & $V_{\text{hel}}$ (km s$^{-1}$) & 203 $\pm$ 0.9 \\
    (10) & $V_{\text{GSR}}$ (km s$^{-1}$) & 113.5 $\pm$ 0.9 \\
    (11) & $\sigma_{v}$ (km s$^{-1}$) & 3.97$^{+0.97}_{-0.86}$ \\
    (12) & Dynamical mass ($M_\odot$) & 3.47$^{+1.6}_{-1.6}$ $\times$ 10$^5$ \\
    (13) & $M/L_V (M_\odot / L_\odot)$ & 2463 $\pm$ 1170 \\
    \bottomrule
    \addlinespace[3pt]
    (14) & Mean [Fe/H] & $-$2.52 $\pm$ 0.10 \\
    (15) & $\sigma_{\mathrm{[Fe/H]}}$ & 0.59 $\pm$ 0.06 \\
\end{tabular}
\smallskip

\footnotesize{\textbf{Notes.} Rows (1)--(2), (4)--(8) are from \cite{munoz_megacam_2018}. Row (3) is from \cite{belokurov_cats_2007}. Rows (9) and (11)--(13) are from \cite{geha_deimos_2026}. Row (10) is from \cite{simon_complete_2011}. Rows (14)--(15) are derived in this paper.}
\end{table}
\vspace{10pt}

Segue~1 has only seven previously published stellar metallicities, which exhibit a trimodal structure, with peaks centered around [Fe/H]~$\approx-1.8,-2.3,$~and~$-3.3$ and an average [Fe/H]~$\approx -2.7$ \citep{norris_extremely_2010, simon_complete_2011, frebel_segue_2014}. If shown to be statistically significant, the apparent peaks in the metallicity distribution are notable given Segue~1’s small stellar mass, as almost all other low-mass UFDs do not exhibit this type of multi-modality in their metallicity distributions \citep{simon_faintest_2019, luna_bimodal_2025}.  

The wide metallicity range in Segue~1 and its extremely low stellar mass make it a natural candidate for the one-shot enrichment scenario described by \citet{frebel_chemical_2012}, in which all stars form in a single, brief burst and are enriched only by the earliest supernovae. Such systems are expected to show large abundance spreads but no evidence for extended chemical evolution. The apparent trimodality of the previous MDF is difficult to reconcile with a strictly one-shot event, as a single burst should not produce multiple distinct metallicity peaks. 

However, \citet{webster_segue_2016} suggested this potential trimodality could be explained by clustered star formation, with Segue~1 hosting evidence for some of the first star-forming clusters. If the galaxy’s small gravitational potential well retained supernova ejecta inefficiently, star formation may have occurred in discrete, chemically distinct bursts rather than in a well-mixed interstellar medium. Since Segue~1 is among the oldest known galaxies, this would be evidence of some of the earliest clustered star formation within a galaxy \citep{brown_quenching_2014}. Alternatively, the observed metallicity distribution may be affected by the small sample size, underscoring the need for additional metallicity measurements. 

\citet{frebel_segue_2014} targeted all seven red giant branch (RGB) member stars, an approach well-suited to obtain full MDFs for classical dwarf spheroidal galaxies where the RGB is sufficiently populated to trace the metallicity distribution (e.g.,~\citealt{, helmi_new_2006, tolstoy_star-formation_2009, kirby_multi-element_2011, pace_2020, tolstoy_2023, walker_2023, barbosa_2025, tolstoy_2025}). In UFDs, however, their small stellar masses yield far fewer RGB stars, limiting the available metallicity measurements. Expanding the sample to include main-sequence turnoff (MSTO) stars is therefore essential for constructing a more complete and statistically robust metallicity distribution for Segue~1.

\begin{figure}[t]
\includegraphics[width=\columnwidth]{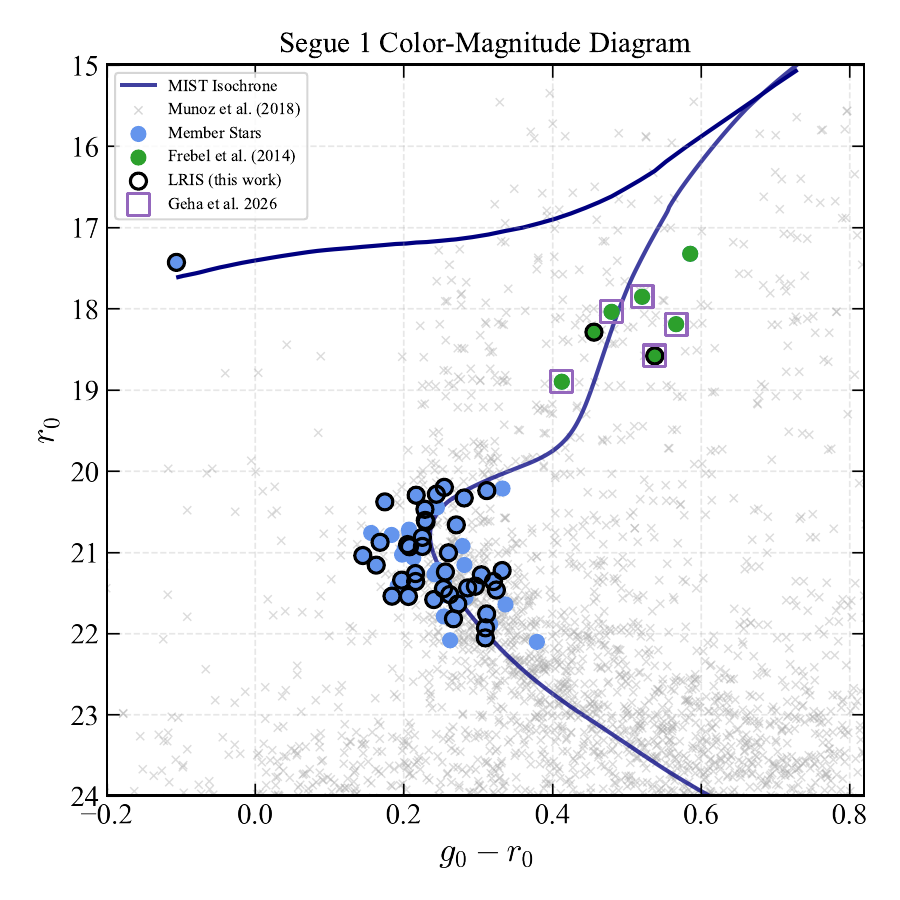}
\caption{Segue~1 color-magnitude diagram. Member stars of Segue~1 from \citet{simon_complete_2011} and \citet{norris_extremely_2010} are shown in blue. \textit{$g_{0}$} and \textit{$r_{0}$} magnitudes are from MegaCam on the Canada--France--Hawaii Telescope \citep{munoz_megacam_2018}. All non-member stars in the Segue~1 field from \citet{munoz_megacam_2018} are shown as gray crosses. Member stars with metallicities derived from calcium triplet (CaT) equivalent widths from \citet{geha_deimos_2026} are indicated by purple squares. Member stars with high-resolution spectroscopy from \citet{frebel_segue_2014} are shown in green. Member stars in this paper with LRIS spectroscopy are shown with a black outline. The dark blue line is an isochrone from MESA Isochrones \& Stellar Tracks (MIST) with a metallicity of [Fe/H] = $-$2.5 and an age of 13.8 Gyr \citep{dotter_mesa_2016, choi_mesa_2016}.}
\label{fig:segue1_cmd}
\end{figure}

By obtaining spectroscopy of the fainter MSTO stars in Segue~1 (Figure~\ref{fig:segue1_cmd}), we significantly increase the metallicity sample size and present an updated MDF based on $N_{\rm stars}=45$, including 40 new measurements from the Low-Resolution Imaging Spectrometer (LRIS; \citealt{oke_keck_1995}, \citealt{mccarthy_blue_1998}). In addition to expanding the sample size, we employ a Gaussian mixture modeling approach to assess whether the MDF of Segue~1 is multimodal or whether the previously observed trimodal structure was an artifact of small-number statistics, and we compare the observed MDF to simple analytical chemical evolution models from \citet{kirby_multi-element_2011} to test whether standard enrichment scenarios can reproduce the metallicity distribution of such a low-mass system. Section~\ref{sec:data} provides an overview of the observations, Section~\ref{sec:methods} describes our metallicity determinations and comparison with previous work, Section~\ref{sec:results} presents the MDF and chemical evolution models, and Section~\ref{sec:discussion} discusses the implications for the star formation history and evolution of Segue~1.

\section{Observations and Data}\label{sec:data}

\begin{table*}[htp]
\renewcommand{\arraystretch}{1.2}
    \centering
    \caption{Summary of Keck/LRIS slit masks used for spectroscopic observations of Segue~1.}
    \label{tab:segue1_slitmasks}
    \begin{tabular}{l c c c c c c}
        \toprule
        Mask Name & $\alpha$ (J2000) & $\delta$ (J2000) & P.A. (deg) & $t_{\text{exp}}$ (s) & MJD & No. of Slits  \\
        \midrule
        s1m1 & 10:07:59.83 & 15:58:54.64 & 75.0 & 28800 & 57777.6 & 25 \\
        s1m3 & 10:08:20.08 & 15:57:08.24 & 66.0 & 11140 & 60021.3 & 25  \\
        s1m4 & 10:07:55.07 & 15:56:57.19 & 143.0 & 4800 & 60022.4 & 16  \\
        \bottomrule
    \end{tabular}
    \vspace{3pt}
    \tablecomments{Columns list the mask name, central right ascension ($\alpha$) and declination ($\delta$) in J2000 coordinates, position angle (P.A.) in degrees, total exposure time ($t_{\text{exp}}$) in seconds, Modified Julian Date (MJD) of observation, and the number of slits on each mask. One slit on mask s1m4 could not be extracted because the Ca~II~K feature was not on the detector.}
\end{table*}

We observed 40 member stars with Ca~II K absorption features in Segue~1 with LRIS on Keck I on 2017 January 23--24 (PI: J. Bland-Hawthorn) and 2023 March 16--18 (PI: A. Ji). We used 1\farcs0-wide slits for all of the targets and used the Atmospheric Dispersion Corrector. For the blue channel of LRIS, we utilized the 600/4000 grism, which provides a spectral resolution of FWHM $\sim4$~\AA. The LRIS spectroscopy with the blue channel covers the wavelength range of $\lambda \sim 3000-5500~ $\AA. The observations from 2017 had seeing ranging from 1\farcs3 to 2\farcs0. In 2023, the seeing ranged from 0\farcs8 to 1\farcs0. While red channel data were also gathered, we only present results based on the blue channel data. Due to some repeated targets across the observations in 2017 and 2023, the total exposure time per target varied between 1.3 h and 11.1 h, with a median of 8.0 h. Slit mask information for these observations can be seen in Table~\ref{tab:segue1_slitmasks}. 

\subsection{Data Reduction}

We used the \texttt{PypeIt} data reduction pipeline \citep{pypeit:joss_pub} to reduce the Keck/LRIS spectra of stars in Segue~1. \texttt{PypeIt} performs the standard calibration procedures, including flat-fielding, wavelength calibration, one-dimensional spectral extraction, and heliocentric velocity correction. Prior to spectral extraction, slit edge tracing and object finding were manually verified to ensure accurate stellar extractions. Wavelength solutions were obtained using arc lamps containing Hg, Ne, Ar, Cd, Zn, Kr, and Xe.

Individual one-dimensional stellar spectra were coadded using \texttt{PypeIt}. After coadding, each spectrum was normalized to the continuum using a polynomial fit excluding strong absorption features.

Stellar radial velocities were determined through cross-correlation of each normalized spectrum with a smoothed, high-resolution rest-frame spectrum of HD~21581, an RGB star with [Fe/H]~$= -1.82$ and a heliocentric velocity of 153.7~km~s$^{-1}$ \citep{roederer_search_2014}. The usable overlapping spectral range between the LRIS data and the template spectrum is 3800--5500~\AA. The low resolution of the LRIS data, combined with flexure effects, results in velocity errors exceeding 10~km~s$^{-1}$ for our spectra. However, the stars in this work are confirmed members according to \citet{simon_complete_2011}, through a spectroscopic survey with the DEep Imaging Multi-Object Spectrograph (DEIMOS, \citealt{faber_deimos_2003}) on Keck II that is complete within 2.3 half-light radii of Segue~1 for stars with colors and magnitudes consistent with membership, though the full sample extends to larger radii. Membership was determined using a combination of photometric, kinematic, and metallicity criteria, requiring stars to have radial velocities consistent with Segue~1's systemic velocity and, where measurable, metallicities consistent with the system. From this analysis, \citet{simon_complete_2011} identify 71 probable members, of which our targets are a subset. Of the 40 stars in our LRIS Segue~1 MDF sample, 12 have proper motion measurements from Gaia DR3; all 12 are consistent with the systemic proper motion of Segue~1 reported by \citet{pace_dwarf_2022}, providing additional kinematic confirmation of their membership.

\section{Methods} \label{sec:methods}
\subsection{Metallicity Determination}\label{sec:metallicity_det} 

To determine stellar metallicities, we employed the empirical calibration developed by \citet{beers_estimation_1999}, which uses a combination of photometric color information and spectroscopic measurements of the Ca II K line strength. We measured the pseudo-equivalent width of the Ca II K absorption feature, known as the KP index. The KP index provides a robust and consistent metric for metallicity-sensitive absorption strength across a wide range of spectral types. The KP index was calculated by integrating the normalized flux across a narrow wavelength band centered on the Ca II K line at 3933.7~\AA, using the continuum windows and line bandpasses defined in \citet{beers_estimation_1999}. The pseudo-equivalent widths are measured over three increasingly broad bandpasses, named K6, K12, and K18, centered on the Ca~II~K line. The final KP value was determined following these criteria:

\[
\text{KP} =
\left\{
\begin{array}{ll}
  \text{K6 ~(3930.7--3936.7\AA)} & \text{if K6} \leq 2\text{\AA}, \\
  \text{K12 (3927.7--3939.7\AA)} & \text{if K6} > 2\text{\AA} \text{, K12} \leq 5\text{\AA}, \\
  \text{K18 (3924.7--3942.7\AA)} & \text{if K12} > 5\text{\AA}.
\end{array}
\right.
\]

\begin{figure*}[ht!]
\plotone{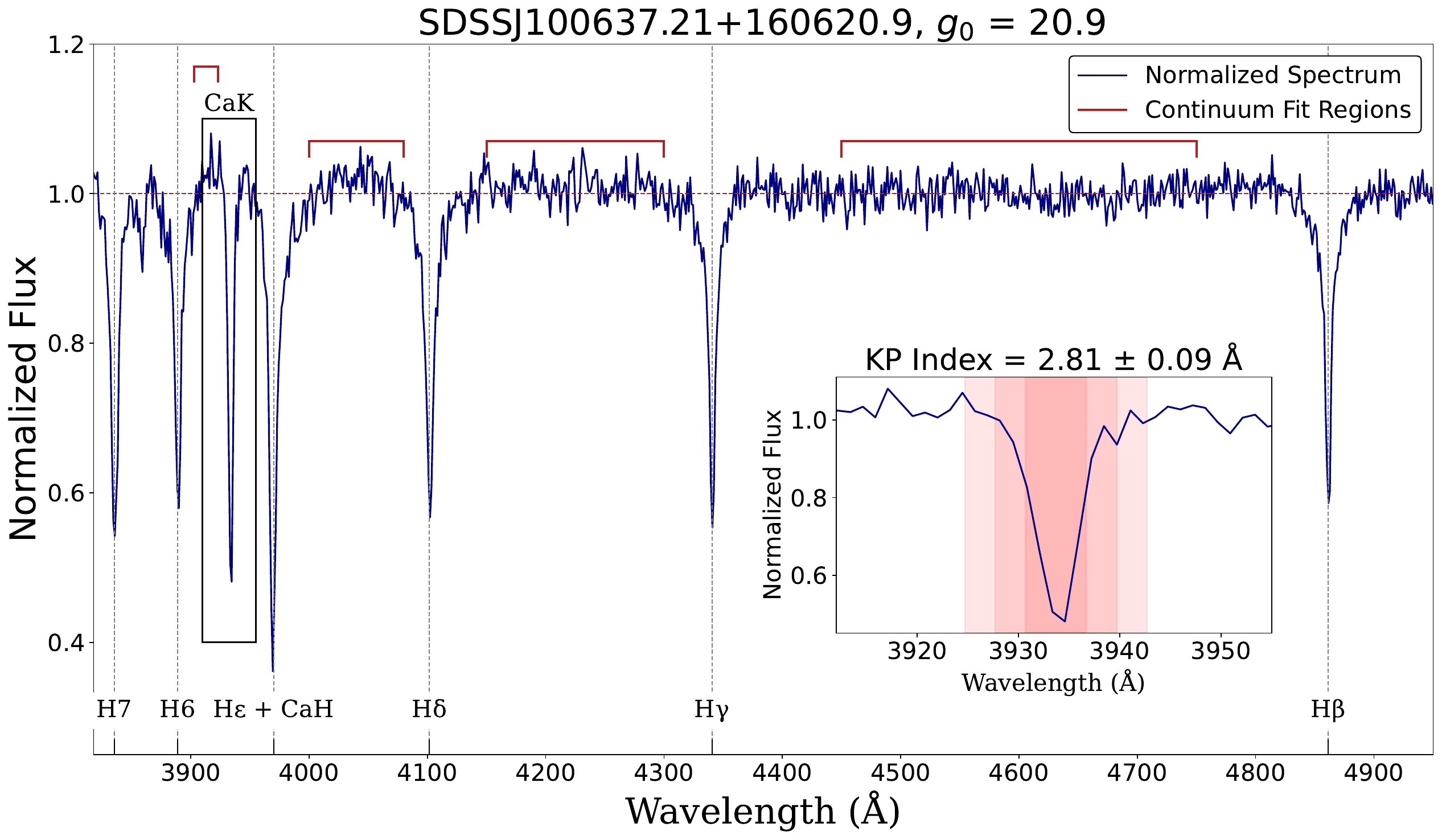}
\caption{Keck/LRIS spectrum of SDSSJ100637.21+160620.9. The spectrum has a signal-to-noise ratio of 54 per pixel in the CaHK region (3900--4000~\AA). Prominent Balmer absorption lines (H7, H6, H$\varepsilon$ + CaH, H$\delta$, H$\gamma$, H$\beta$) are labeled. The red brackets indicate the continuum regions used for normalization. The Ca~II K line at 3933~\AA\ is highlighted on the spectrum in the black box. The inset black box shows a zoom-in of the Ca~II K feature with K12 index, with the shaded red regions corresponding to the $K6$, $K12$, and $K18$ bandpasses used to compute the KP index, following the method of \citet{beers_estimation_1999}.} \label{fig:spectrum}
\end{figure*}

Figure~\ref{fig:spectrum} illustrates the Ca~II~K region used in this measurement, showing all three bandpasses. At the wavelength of Ca~II~K (3933~\AA), the signal-to-noise ratio (S/N) of the final coadded spectra peaked at S/N~$\approx$~530 for the brightest target at $g_{\rm 0} = 17.3$, decreasing to S/N~$\approx$~60 at $g_{\rm 0} = 20.6$, and S/N~$\approx$~19 at $g_{\rm 0} = 22.3$. An example of a reduced and normalized one-dimensional spectrum is shown in Figure~\ref{fig:spectrum}.

Color information for the stellar sample was derived from MegaCam $g$ and $r$ photometry \citep{munoz_megacam_2018}. We compared available photometric sources including SDSS, MegaCam, and DELVE, finding that MegaCam provided the lowest photometric uncertainties and deepest coverage for this work. We adopt the MegaCam photometry from \citet{munoz_survey_2018}, which is calibrated to the SDSS photometric system and corrected for extinction as described in their Paper~I. We then converted these magnitudes to Johnson-Cousins $B-V$ colors using the transformations from \citet{jordi_empirical_2006}.

Using the derived $B-V$ color and KP index, we determined [Fe/H] metallicities by applying the \citet{beers_estimation_1999} empirical calibration. The resulting [Fe/H] values provide a robust estimate of each star’s metallicity, with typical uncertainties of $\sim0.25$–$0.4$ dex. These uncertainties, listed in Table~\ref{tab:segue1_fehs}, arise from the MegaCam \textit{g}- and \textit{r}-band photometric errors propagated in quadrature through the $g-r \rightarrow B-V$ color conversion ($\langle \sigma_{\rm phot} \rangle = 0.03$), the spectroscopic uncertainty from the KP index ($\langle \sigma_{\rm spec} \rangle = 0.05$), and the uncertainty in the empirical calibration ($\langle \sigma_{\rm calib} \rangle = 0.28$). Additionally, we impose minimum floors of 0.02 dex on both $\sigma_{\rm phot}$ and $\sigma_{\rm spec}$, reflecting the minimum sensitivity of the calibration to small perturbations in KP~index and $B-V$. The overall uncertainty on our metallicity measurements is $\langle \sigma_{\rm total} \rangle = 0.30$. 

Two stars in our LRIS sample also have CaT metallicities from \citet{geha_deimos_2026}, providing a direct external check on our Ca II K measurements: SDSSJ100652.33+160235.8 and SDSSJ100710.08+160623.9. For both stars, our LRIS Ca II K metallicities agree with the CaT values within the uncertainties: for SDSSJ100710.08+160623.9, $\Delta\text{[Fe/H]} = +0.27$ (CaK $-0.95 \pm 0.62$, CaT $-1.22 \pm 0.17$), and for SDSSJ100652.33+160235.8, $\Delta\text{[Fe/H]} = +0.30$ (CaK $-3.11 \pm 0.32$, CaT $-3.41 \pm 0.19$), suggesting that the two scales are broadly consistent to within $\sim 0.3$ dex. These same stars also appear in \citet{frebel_segue_2014}, whose high-resolution abundances are $\approx$~0.6 dex lower than the LRIS metallicities and $\approx$~0.32 dex lower than the CaT metallicities. Notably, both stars show standard $\alpha$-enhanced [Ca/Fe]~$\approx+0.57$ in \citet{frebel_segue_2014}, consistent with the rest of the Segue~1 sample, suggesting that the offset reflects a systematic scale difference rather than anomalous abundances in these two stars. The CaT scale is preferred because it is tied to the standard mass-metallicity relations used in the literature \citep{carrera_near-infrared_2013, navabi_revisiting_2026}. To place the \citet{frebel_segue_2014} metallicities on this scale, we apply a +0.32 dex shift, equal to the mean offset between \citet{frebel_segue_2014} and the CaT metallicities.

\subsection{MDF Modeling} \label{sec:modeling}

Given the presence of multiple modes in the MDF in \citet{frebel_segue_2014}, we compare single and multimodal Gaussian models to determine the best representation of the data. A Gaussian MDF, defined by its mean metallicity ($\mu_m$) and dispersion ($\sigma_{m}$), can be generalized to a Gaussian mixture model (GMM) with $M$ components defined as

{\small
\begin{equation}
p([\mathrm{Fe/H}]) = 
\sum_{m=1}^{M} w_m 
\frac{1}{\sigma_m\sqrt{2\pi}}
\exp\!\left[-\frac{([\mathrm{Fe/H}] - \mu_m)^2}{2\sigma_m^2}\right],
\end{equation}
}

\noindent where each component $m$ is described by a mean metallicity $\mu_m$, metallicity dispersion $\sigma_m$, and weight $w_m$ such that $\sum_{m=1}^{M} w_m = 1$. The case $M=1$ corresponds to the single-Gaussian MDF, while $M=2$ and $M=3$ describe bimodal and trimodal models, respectively. 

To fit the MDF, we maximize the log-likelihood function

\begin{align}
&\ln \mathcal{L}(\{\mu_m, \sigma_m, w_m\}) = \notag \\[6pt]
&\sum_{i=1}^{N} \ln \left(
    \sum_{m=1}^{M} w_m \,
    \mathcal{N}\!\left([{\rm Fe/H}]_i \,\middle|\,
    \mu_m, \sqrt{\sigma_m^{2} + \sigma_i^{2}}\right)
  \right),
\end{align}

\noindent where $\mathcal{N}([{\rm Fe/H}]_i \,|\, \mu_m, \sigma_m)$ is the Gaussian probability density for star $i$. Measurement uncertainties are incorporated by convolving each Gaussian with the individual metallicity error $\sigma_i$ of the star. 

Model selection is performed using the Akaike Information Criterion (AIC), 
\begin{equation}
{\rm AIC} = 2k - 2\ln \mathcal{L},
\end{equation}
where $k$ is the number of free parameters \citep{akaike_new_1974}. The model with the lowest AIC value is statistically preferred, balancing goodness-of-fit against model complexity. We use the AIC rather than the Bayesian Information Criterion (BIC, \citealt{schwarz_1978}) because the AIC penalizes extra parameters less, making it more likely to favor multi-component models. A preference for the single Gaussian under the AIC is therefore a conservative result.

Additionally, we compare the observed metallicity distribution of Segue~1 to predictions from chemical evolution models\footnote{ \url{https://github.com/alexji/mdfmodels}.}. Three models are analyzed in this paper: leaky box, pre-enriched box, and extra gas \citep{kirby_multi-element_2011}. We fit these analytic models to the observed MDF using a maximum-likelihood framework, following the formalism of \citet{kirby_multi-element_2011}. The leaky box assumes a system with a finite supply of pristine gas ([Fe/H]$_0 = -\infty$), no accretion, and metal loss governed by an effective yield $p$ \citep{pagel_nucleosynthesis_1997}. The pre-enriched model allows for a nonzero initial metallicity. The extra gas model includes accretion of pristine gas, with the infall parameter $M_{\rm infall}$ \citep{lynden-bell_chemical_1975}. 

\section{Results} \label{sec:results}

\begin{table*}[!htbp]
\raggedright
\begingroup
\scriptsize
\setlength{\tabcolsep}{3.5pt}
\caption{Photometric and Spectroscopic Properties of Segue~1 Member Stars.}
\label{tab:segue1_fehs}
\begin{tabular}{cccccccccccccc}
\toprule
Star Name & Masks & \textit{$g_{0}$} & \textit{$r_{0}$} & $B\!-\!V$ & Mem. & S/N & KP Index (\AA) & [Fe/H] & $\sigma_{\rm spec}$ & $\sigma_{\rm phot}$ & $\sigma_{\rm calib}$ & $\sigma_{[\mathrm{Fe/H}]}$ & Comment \\
\midrule
SDSSJ100629.08+160623.8 & s1m4 & 21.58 & 21.35 & 0.44 & 1 & 31 & 3.10\,$\pm$\,0.14 & $-2.10$ & 0.03 & 0.12 & 0.25 & 0.28 &  \\
SDSSJ100631.84+160626.3 & s1m4 & 21.62 & 21.37 & 0.45 & 1 & 31 & 1.26\,$\pm$\,0.11 & $-2.99$ & 0.10 & 0.03 & 0.30 & 0.32 &  \\
SDSSJ100633.35+160327.4 & s1m4 & 22.14 & 21.83 & 0.50 & 1 & 23 & 1.84\,$\pm$\,0.14 & $-2.75$ & 0.07 & 0.05 & 0.30 & 0.31 &  \\
SDSSJ100636.09+160628.8 & s1m4 & 21.85 & 21.48 & 0.55 & 1 & 25 & 5.42\,$\pm$\,0.23 & $-2.03$ & 0.07 & 0.05 & 0.23 & 0.25 &  \\
SDSSJ100636.26+160142.0 & s1m4 & 20.59 & 20.39 & 0.42 & 1 & 65 & 0.62\,$\pm$\,0.04 & $-3.56$ & 0.02 & 0.03 & 0.32 & 0.32 &  \\
SDSSJ100637.21+160620.9 & s1m4 & 20.90 & 20.64 & 0.47 & 1 & 54 & 2.81\,$\pm$\,0.09 & $-2.27$ & 0.03 & 0.04 & 0.26 & 0.26 &  \\
SDSSJ100638.75+160613.7 & s1m4 & 21.76 & 21.55 & 0.43 & 1 & 30 & 1.50\,$\pm$\,0.11 & $-2.75$ & 0.09 & 0.04 & 0.29 & 0.30 &  \\
SDSSJ100640.46+160238.1 & s1m4 & 21.84 & 21.53 & 0.50 & 1 & -- & -- & -- & -- & -- & -- & -- & (a) \\
SDSSJ100643.47+160604.5 & s1m4 & 21.08 & 20.89 & 0.41 & 1 & 44 & 0.47\,$\pm$\,0.08 & $-3.94$ & 0.19 & 0.03 & 0.32 & 0.37 &  \\
SDSSJ100644.55+160129.4 & s1m4 & 20.56 & 20.31 & 0.46 & 1 & 66 & 0.95\,$\pm$\,0.06 & $-3.42$ & 0.11 & 0.01 & 0.32 & 0.34 &  \\
SDSSJ100649.05+160348.7 & s1m1; s1m4 & 20.75 & 20.48 & 0.47 & 1 & 75 & 3.71\,$\pm$\,0.05 & $-2.11$ & 0.03 & 0.01 & 0.24 & 0.24 &  \\
SDSSJ100649.63+160308.3 & s1m3; s1m4 & 21.09 & 20.83 & 0.46 & 1 & 60 & 1.86\,$\pm$\,0.05 & $-2.66$ & 0.02 & 0.01 & 0.29 & 0.29 &  \\
SDSSJ100652.23+160349.2 & s1m4 & 21.55 & 21.26 & 0.49 & 1 & 41 & 3.31\,$\pm$\,0.09 & $-2.25$ & 0.02 & 0.01 & 0.25 & 0.25 &  \\
SDSSJ100652.33+160235.8 & s1m4 & 18.83 & 18.31 & 0.67 & 1 & 183 & 1.90\,$\pm$\,0.02 & $-3.11$ & 0.02 & 0.05 & 0.32 & 0.32 & (b) \\
SDSSJ100654.46+160126.6 & s1m1 & 21.36 & 21.17 & 0.41 & 1 & 33 & 4.30\,$\pm$\,0.13 & $-1.63$ & 0.04 & 0.11 & 0.23 & 0.26 &  \\
SDSSJ100655.45+160416.2 & s1m1; s1m3 & 20.67 & 20.35 & 0.51 & 1 & 51 & 5.54\,$\pm$\,0.09 & $-1.75$ & 0.04 & 0.01 & 0.23 & 0.23 &  \\
SDSSJ100657.42+160300.0 & s1m1 & 21.62 & 21.24 & 0.56 & 1 & 25 & 6.15\,$\pm$\,0.20 & $-1.82$ & 0.07 & 0.10 & 0.26 & 0.29 &  \\
SDSSJ100657.62+160230.1 & s1m3 & 22.43 & 22.07 & 0.54 & 1 & 27 & 3.44\,$\pm$\,0.18 & $-2.39$ & 0.02 & 0.07 & 0.26 & 0.26 &  \\
SDSSJ100659.10+160437.0 & s1m1; s1m3 & 21.52 & 21.27 & 0.45 & 1 & 30 & 1.18\,$\pm$\,0.09 & $-3.19$ & 0.10 & 0.02 & 0.31 & 0.33 &  \\
SDSSJ100659.76+160218.5 & s1m1; s1m3 & 21.75 & 21.46 & 0.49 & 1 & 42 & 1.95\,$\pm$\,0.10 & $-2.71$ & 0.06 & 0.01 & 0.29 & 0.30 &  \\
SDSSJ100700.72+160609.8 & s1m3 & 21.97 & 21.65 & 0.50 & 1 & 19 & 0.59\,$\pm$\,0.18 & $-3.70$ & 0.30 & 0.01 & 0.34 & 0.45 &  \\
SDSSJ100700.90+160400.8 & s1m3 & 21.87 & 21.60 & 0.47 & 1 & 27 & 0.80\,$\pm$\,0.14 & $-3.42$ & 0.12 & 0.01 & 0.32 & 0.34 &  \\
SDSSJ100701.34+160200.0 & s1m1; s1m3 & 20.51 & 20.21 & 0.49 & 1 & 63 & 1.80\,$\pm$\,0.04 & $-2.71$ & 0.02 & 0.01 & 0.29 & 0.29 &  \\
SDSSJ100701.40+160440.6 & s1m1; s1m3 & 21.18 & 20.94 & 0.44 & 1 & 38 & 1.19\,$\pm$\,0.07 & $-3.15$ & 0.10 & 0.01 & 0.30 & 0.32 &  \\
SDSSJ100703.01+160425.0 & s1m1; s1m3 & 20.88 & 20.62 & 0.46 & 1 & 52 & 0.82\,$\pm$\,0.05 & $-3.42$ & 0.02 & 0.01 & 0.32 & 0.32 &  \\
SDSSJ100703.15+160335.0 & s1m1 & 21.79 & 21.56 & 0.44 & 1 & 24 & 4.57\,$\pm$\,0.15 & $-1.72$ & 0.04 & 0.16 & 0.22 & 0.28 &  \\
SDSSJ100703.26+160234.4 & s1m1 & 20.61 & 20.26 & 0.54 & 1 & 48 & 4.34\,$\pm$\,0.07 & $-2.23$ & 0.02 & 0.01 & 0.23 & 0.23 &  \\
SDSSJ100703.32+160140.6 & s1m1; s1m3 & 21.15 & 20.92 & 0.44 & 1 & 55 & 2.93\,$\pm$\,0.15 & $-2.17$ & 0.03 & 0.05 & 0.25 & 0.26 &  \\
SDSSJ100704.35+160459.4 & s1m1 & 21.32 & 21.02 & 0.49 & 1 & 36 & 4.07\,$\pm$\,0.10 & $-2.10$ & 0.03 & 0.01 & 0.23 & 0.23 &  \\
SDSSJ100705.60+160422.0 & s1m1 & 17.31 & 17.43 & 0.16 & 1 & 531 & 0.80\,$\pm$\,0.01 & $-2.90$ & 0.02 & 0.01 & 0.29 & 0.29 &  \\
SDSSJ100706.75+160444.4 & s1m1; s1m3 & 21.77 & 21.43 & 0.53 & 1 & 31 & 6.37\,$\pm$\,0.13 & $-1.63$ & 0.04 & 0.06 & 0.26 & 0.27 &  \\
SDSSJ100708.51+160401.5 & s1m3 & 21.22 & 21.05 & 0.39 & 1 & 42 & 3.52\,$\pm$\,0.12 & $-1.64$ & 0.04 & 0.01 & 0.24 & 0.25 &  \\
SDSSJ100708.92+160445.6 & s1m1; s1m3 & 21.78 & 21.45 & 0.52 & 1 & 57 & 3.74\,$\pm$\,0.07 & $-2.28$ & 0.02 & 0.07 & 0.24 & 0.26 &  \\
SDSSJ100709.94+160503.2 & s1m1; s1m3 & 22.30 & 21.95 & 0.54 & 1 & 19 & 1.48\,$\pm$\,0.12 & $-3.10$ & 0.08 & 0.01 & 0.32 & 0.33 &  \\
SDSSJ100710.08+160623.9 & s1m1; s1m3 & 19.22 & 18.60 & 0.75 & 1 & 123 & 10.41\,$\pm$\,0.04 & $-0.95$ & 0.02 & 0.01 & 0.62 & 0.62 & (c) \\
SDSSJ100710.27+160355.9 & s1m1 & 22.13 & 21.77 & 0.54 & 1 & 19 & 4.45\,$\pm$\,0.19 & $-2.23$ & 0.03 & 0.08 & 0.23 & 0.25 &  \\
SDSSJ100714.94+160448.8 & s1m1; s1m3 & 20.99 & 20.68 & 0.50 & 1 & 53 & 1.37\,$\pm$\,0.05 & $-3.07$ & 0.02 & 0.02 & 0.31 & 0.32 &  \\
SDSSJ100715.09+160708.2 & s1m3 & 21.64 & 21.29 & 0.53 & 1 & 30 & 3.18\,$\pm$\,0.16 & $-2.43$ & 0.02 & 0.01 & 0.26 & 0.26 &  \\
SDSSJ100715.72+160260.0 & s1m1; s1m3 & 21.20 & 20.94 & 0.46 & 1 & 49 & 3.20\,$\pm$\,0.09 & $-2.21$ & 0.03 & 0.01 & 0.25 & 0.25 &  \\
SDSSJ100715.80+160544.2 & s1m1; s1m3 & 20.58 & 20.30 & 0.48 & 1 & 78 & 1.85\,$\pm$\,0.03 & $-2.71$ & 0.02 & 0.03 & 0.29 & 0.29 &  \\
SDSSJ100716.26+160340.3 & s1m3 & 21.74 & 21.38 & 0.55 & 1 & 23 & 4.37\,$\pm$\,0.19 & $-2.23$ & 0.05 & 0.04 & 0.23 & 0.24 &  \\
SDSSJ100650.83+160351.2 & s1m4 & 22.19 & 21.86 & 0.52 & 0 & 24 & 1.15\,$\pm$\,0.09 & $-3.26$ & 0.07 & 0.02 & 0.32 & 0.33 &  \\
SDSSJ100713.69+160444.8 & s1m3 & 21.95 & 21.54 & 0.58 & 0 & 21 & 6.72\,$\pm$\,0.31 & $-1.77$ & 0.13 & 0.06 & 0.29 & 0.32 &  \\
SDSSJ100717.36+160355.6 & s1m1; s1m3 & 20.02 & 19.67 & 0.53 & 0 & 102 & 5.88\,$\pm$\,0.04 & $-1.79$ & 0.02 & 0.01 & 0.24 & 0.24 &  \\
\bottomrule
\end{tabular}
\endgroup
\vspace{3pt}
\tablecomments{Listed columns are: star name, observing mask(s), dereddened $g$ and $r$ magnitudes, $B\!-\!V$ color, membership from \cite{simon_complete_2011}, signal-to-noise ratio per pixel in the CaHK region (3900--4000~\AA), KP index in \AA, derived [Fe/H], and the associated uncertainties ($\sigma_{\rm spec}$ from spectroscopic errors, $\sigma_{\rm phot}$ from photometric color, $\sigma_{\rm calib}$ from calibration uncertainty, and the total $\sigma_{[\mathrm{Fe/H}]}$).\\
(a) Ca~II~K not on detector.\\
(b) Frebel et al. (2014): [Fe/H] = $-3.60$.\\
(c) Frebel et al. (2014): [Fe/H] = $-1.67$.}
\end{table*}

\begin{figure}
    \centering
    \includegraphics[width=\columnwidth]{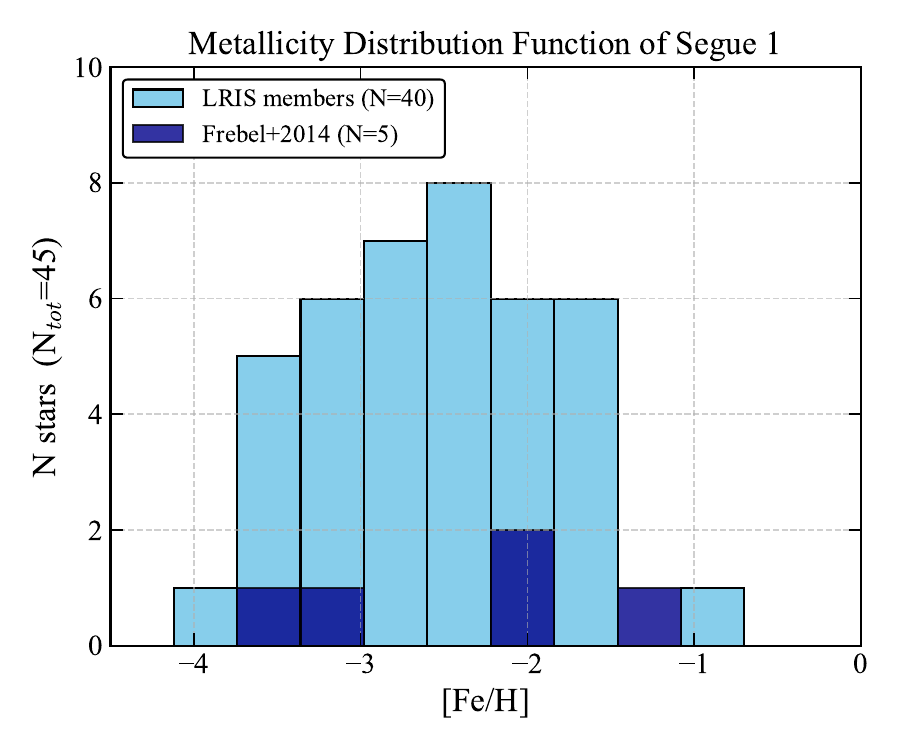}
    \caption{Metallicity distribution function (MDF) of Segue~1. The cyan histogram shows the MDF from this work, including LRIS spectroscopy of 40 member stars, with bin widths of 0.4 dex. The dark blue histogram represents the MDF from \citet{frebel_segue_2014}, offset by +0.32 dex. A single Gaussian fit gives an intrinsic metallicity dispersion of $\sigma_{\mathrm{int}} = 0.59 \pm 0.06 \,\mathrm{dex}$, with a mean metallicity of $\mu = -2.52 \pm 0.10 \,\mathrm{dex}$. Our expanded sample reveals a broader and more continuous MDF compared to the discrete peaks seen in the previous study.}
    \label{fig:combined_mdf}
\end{figure}

Table~\ref{tab:segue1_fehs} and Figure~\ref{fig:combined_mdf} include measured [Fe/H] values for Segue~1 stars from LRIS observations. The best-fit mean metallicity and intrinsic dispersion were determined by maximizing the Gaussian log-likelihood function, where the likelihood incorporates the individual [Fe/H] uncertainties to distinguish intrinsic scatter from observational error. We find a mean metallicity of [Fe/H] $= -2.52 \pm 0.10$ with an intrinsic dispersion of $\sigma_{\mathrm{[Fe/H]},\rm int} = 0.59 \pm 0.06$ with no evidence of distinct peaks. The metallicities range from [Fe/H] $\approx -3.9$ to [Fe/H] $\approx -0.95$. 
This MDF represents a six-fold increase in the sample size relative to \citet{frebel_segue_2014}, with $N_{\rm tot} = 45$ stars. 

\begin{figure*}[!htbp]
    \centering
    \vspace{-0.5em}
    \includegraphics[width=\textwidth]{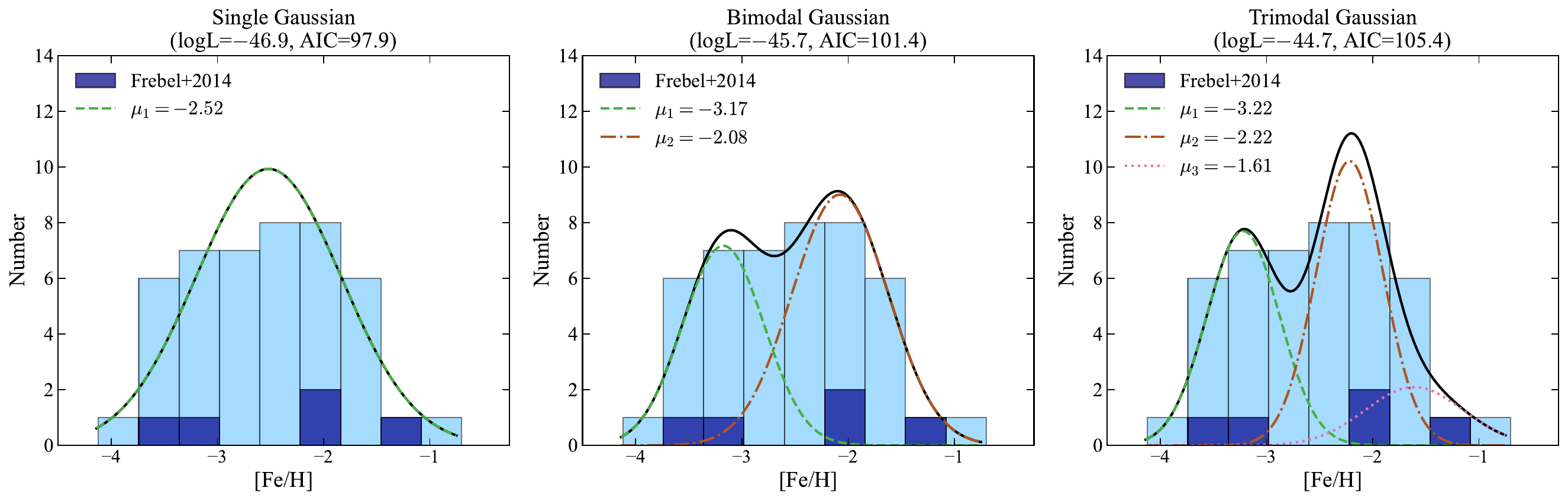}
    \caption{Comparison of model metallicity distribution functions (MDFs) for Segue~1. Metallicities from LRIS spectroscopy are shown in light blue and the \citet{frebel_segue_2014} metallicities are shown in dark blue (and offset by +0.32 dex). The left panel shows a single Gaussian fit to the MDF, with a mean metallicity of $\mu = -2.52$ and an intrinsic dispersion of $\sigma_{\mathrm{int}} = 0.59$. The middle panel displays a bimodal Gaussian mixture model, with components at $\mu = -2.08$ and $\mu = -3.17$. The right panel shows a trimodal Gaussian mixture model, with components at $\mu = -3.22$, $-2.22$, and $-1.61$. The colored dashed lines represent the individual Gaussian components, while the solid black line shows the combined mixture model in each case. Models have been convolved with a typical 0.30 dex uncertainty for display, but the likelihood fits use the star-by-star heteroscedastic uncertainties. The AIC prefers the single Gaussian (Table 4).}
    \label{fig:mdf_models}
    \vspace{-0.5em}
\end{figure*}

Table~\ref{tab:mdf_fits} and Figure~\ref{fig:mdf_models} show the results of the multi-component Gaussian analysis: while the maximum log likelihood improves with each additional Gaussian component, the AIC penalizes the increasing number of free parameters. The single Gaussian model has the lowest AIC value (AIC = 97.9, logL = $-46.9$), compared to the bimodal (AIC = 101.4, logL = $-45.7$) and trimodal (AIC = 105.4, logL = $-44.7$) fits. Thus, although the trimodal model achieves a better log-likelihood, its $\Delta$AIC $= +$7.5 relative to the single Gaussian indicates the extra parameters do not significantly improve the fit.

\begin{figure}[ht]
    \centering
    \includegraphics[width=\columnwidth]{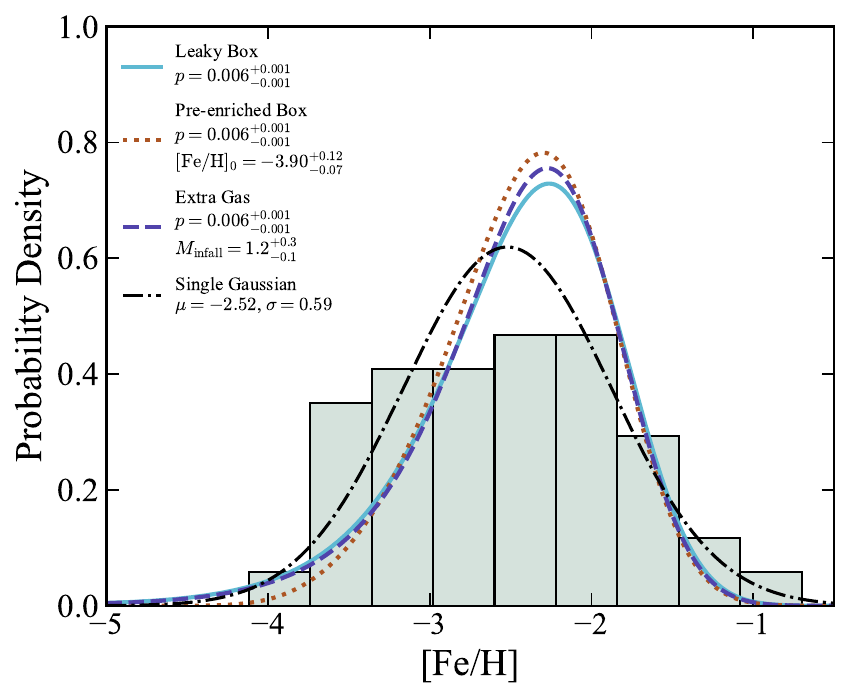}
    \caption{Metallicity distribution function (MDF) of Segue~1 compared with simple analytical chemical evolution models. The models—leaky box, extra gas, and pre-enriched box—are convolved with a representative observational uncertainty of 0.30 dex to match the resolution of the data. The best-fit single Gaussian is shown for comparison (dash-dotted line).}
    \label{fig:chemev_models}
\end{figure}

Table~\ref{tab:mdf_fits} lists the best-fit parameters and AIC values for the chemical-evolution models alongside the single-Gaussian MDF fit shown in Figure~\ref{fig:mdf_models}. None of the analytic models—leaky box, pre-enriched, or extra gas—are statistically preferred over the Gaussian. Figure~\ref{fig:chemev_models} provides a visual comparison of these chemical-evolution model predictions with the observed MDF, after convolution with a representative observational uncertainty of 0.29 dex. This comparison highlights that the analytic models do not capture additional structure beyond what is already described by a single Gaussian. In particular, the MDF of Segue~1 does not exhibit the narrow peak of these analytic models and the extended metal-poor tail commonly seen in more massive dwarf galaxies \citep{helmi_new_2006, kirby_universal_2013}. Together with the mixture-model results, this indicates that Segue~1 occupies a chemically distinct regime relative to more massive systems, though the detailed form of its MDF may be limited by the current sample size.

\begin{table*}[t]
\renewcommand{\arraystretch}{1.3}
    \centering
    \caption{Best-fit parameters for the Segue~1 MDF with analytical chemical evolution models.}
    \label{tab:mdf_fits}
    \begin{tabular}{lcccc}
        \toprule
        Model & log(p) & Parameters & Number of Parameters (k) & AIC \\
        \midrule
        \textbf{Single Gaussian} & --- & \textbf{$\mu = -2.52 \pm 0.10,\ \sigma = 0.59 \pm 0.06 $} & 2 & \textbf{97.9} \\
        Two Gaussians & --- & $\mu_{1} = -3.17,~ \mu_{2} = -2.08 $ & 5 & 101.4 \\
        Three Gaussians & ---
        & $\mu_{1} = -3.22,~ \mu_{2} = -2.22,~ \mu_{3} = -1.61$ & 8 & 105.4 \\
        \midrule
        Leaky Box & $-2.21^{+0.08}_{-0.08} $ & --- & 1 &  98.0 \\
        Pre-Enriched & $-2.26^{+0.09}_{-0.08}$ & [Fe/H]$_0 =-3.90^{+0.12}_{-0.07}$ & 2 & 101.6 \\
        Extra Gas & $-2.22^{+0.08}_{-0.08}$ & $M_{\rm infall} = 1.2^{+0.30}_{-0.10}$ & 2 & 100.0 \\
        \bottomrule
    \end{tabular}
    \vspace{3pt}
    \tablecomments{The number of free parameters $k$ counts the component means, dispersions, and mixing fractions for Gaussian mixtures, and the yield plus any additional model-specific terms (e.g., initial metallicity or infall parameter $M_{\rm infall}$) for the chemical evolution models. }
\end{table*}

\section{Discussion} \label{sec:discussion}
With this work, we expand the spectroscopic sample of Segue~1 from 7 red giant stars \citep{frebel_segue_2014} to 45 member stars, a six-fold increase. This is the most comprehensive metallicity study of Segue~1 to date, enabling a statistically robust characterization of its MDF ($\mu = -2.52 \pm 0.10 \,\mathrm{dex}$, $\sigma_{\mathrm{int}} = 0.59 \pm 0.06 \,\mathrm{dex}$) and comparisons to other UFDs with published MDFs. Additionally, the larger sample allows us to reassess proposed formation scenarios.

\vspace{3pt}

\subsection{Comparison to other UFD metallicity distributions}
This result adds to a growing body of work that characterizes MDFs across the UFD population. In particular, \citet{fu_metallicity_2023} used CaHK and HST photometry to derive homogeneous MDFs for 13 UFDs, finding typical metallicity dispersions of $\sigma_{\mathrm{[Fe/H]}} \approx 0.3-0.7$ dex at $-7 \leq M_V \leq -1$ and showing internal [Fe/H] dispersions are ubiquitous across dwarf galaxies. For Segue~1, they infer $\langle\mathrm{[Fe/H]}\rangle \approx -2.36$ and $\sigma_{\mathrm{[Fe/H]}} \approx 0.65~\mathrm{dex}$ from a sample of 12 stars, broadly consistent with our LRIS-based MDF. For the five stars in common between our sample and \citet{fu_metallicity_2023}, the individual [Fe/H] measurements agree with a mean offset of $\langle\Delta\mathrm{[Fe/H]}\rangle = +0.10 \pm 0.43$ (LRIS~--~Fu), with all five stars consistent within $1.5\sigma$. The inclusion of additional MSTO stars in this work expands the Segue~1 MDF while confirming that Segue~1 lies at the metal-poor, high-dispersion end of UFD properties.

Segue~1's position at the high-dispersion end of the UFD population (\citealt{fu_metallicity_2023, geha_deimos2_2026}; this work) is consistent with an emerging mass-dependent trend at the lowest stellar masses. Spectroscopic samples of DEIMOS-observed dwarfs show that $\sigma_{\rm [Fe/H]}$ is relatively modest ($\sim0.3$~dex) for systems above $\log[M_\star/M_\odot]~\approx 4$, but increases sharply for the lowest-mass systems \citep{geha_deimos2_2026}. Segue~1's large dispersion ($\sigma_{\rm [Fe/H]}=0.59\pm0.06$) sits at the extreme low-mass end and reinforces this upturn --- hinting that something qualitatively different may govern chemical enrichment below this stellar mass threshold. For example, an extreme enrichment history with stochastic and inhomogeneous metal mixing can produce a broad metallicity spread even within a brief star formation episode \citep{frebel_chemical_2012, webster_segue_2016, emerick_simulating_2020}. A definitive characterization of the relation between $\sigma_{\rm [Fe/H]}$ and log$[M_\star/M_\odot]$ at the faint end will nonetheless require larger, homogeneously analyzed UFD samples spanning the full range of stellar masses below $\log[M_\star/M_\odot] \approx 4$. 

While population-wide studies establish broad trends among UFDs, recent detailed MDF measurements of individual systems such as Reticulum~II ($N_{\rm stars} = 167$) and Bo\"{o}tes~I ($N_{\rm stars} = 92$) show MDF morphologies that differ from that of Segue~1. Reticulum~II's spectroscopic MDF (including main-sequence turnoff stars) is bimodal, with $\sim$80\% of stars clustered near ${\rm [Fe/H]}\approx -3.0$ and a second, more metal-rich component near ${\rm [Fe/H]}\approx -2.1$, separated by a pronounced gap interpreted as evidence for a long quiescent period and extended chemical evolution \citep{luna_bimodal_2025}. Early work on Bo\"{o}tes~I based on a smaller sample ($N_{\rm stars} = 41$) found that the same simple chemical-evolution models referenced in Section~\ref{sec:modeling} (e.g., leaky box, pre-enriched box) provide comparably good fits to the MDF, with no model uniquely preferred \citep{lai_feh_2011}. \citet{romano_chemical_2014} extended this analysis with both classical and cosmologically-motivated models, finding that multiple chemical evolution models reproduce the observed MDF reasonably well, and concluding that a larger homogeneous stellar sample is needed to distinguish among them. This inability to distinguish between models was attributed in part to limited sample size, a degeneracy that may similarly apply to Segue~1, where additional MDF substructure could remain unresolved with current data. 

Using VLT spectroscopy, \citet{jenkins_very_2021} rederived the Bo\"{o}tes~I MDF, finding a mean metallicity of $\langle{\rm [Fe/H]}\rangle = -2.34$ and an intrinsic dispersion of $\sigma_{\mathrm{[Fe/H]}} = 0.28$~dex. \citet{jenkins_very_2021} also found that the leaky box model is disfavored compared to the pre-enriched and extra gas models. More recently, \citet{sandford_chemodynamics_2025} analyzed an expanded spectroscopic sample ($N_{\rm stars} = 92$) and recovered a similar mean metallicity ($\langle{\rm [Fe/H]}\rangle = -2.43$) and dispersion ($\sigma_{\rm [Fe/H]} = 0.28$~dex). The bimodality seen in Reticulum~II and the comparatively narrow dispersion of Bo\"{o}tes~I, relative to Segue~1, show that the Segue~1 MDF is distinct from other well-studied UFDs, despite similar, metal-poor mean metallicities.

\subsection{Galaxy Formation Scenarios for Segue~1}

Several formation pathways have been proposed for Segue~1. First, a simple, continuous star formation history (SFH) would be expected to produce an MDF with a single dominant component and a metal-poor tail, as is typical for the MDFs of more massive, classical dwarf galaxies and those described by the chemical evolution models discussed in Section~\ref{sec:modeling} \citep{kirby_multi-element_2011}. Second, a one-shot enrichment scenario first described by \citet{frebel_chemical_2012} and later applied to Segue~1 by \citet{frebel_segue_2014} suggests that Segue~1 formed all of its stars in a single, brief burst, with chemical signatures reflecting only the earliest supernovae. The resulting spread in [Fe/H] in this model arises from inhomogeneous mixing of supernova ejecta, leading to locally enriched gas pockets from which stars of different metallicities formed. Finally, multiple discrete bursts of clustered star formation---potentially producing a multimodal MDF---have also been proposed for Segue~1 by \citet{webster_segue_2016}, motivated in part by the apparent three-peaked distribution observed in the 7-star sample of \citet{frebel_segue_2014}. In the remainder of this section, we consider each of these three scenarios in turn.

For the continuous SFH case, our newly derived MDF is consistent with a single peak and is well described by a single Gaussian with $\sigma_{\mathrm{[Fe/H]}} = 0.59 \pm 0.06$ dex. While this behavior is similar to that reported for many UFDs, the apparent Gaussianity—both in Segue~1 and in other systems—likely reflects the limited size of existing spectroscopic samples and, in some cases, large measurement uncertainties, rather than an intrinsically Gaussian MDF \citep[e.g.,][]{lai_feh_2011, fu_metallicity_2023, luna_bimodal_2025}. In this low-number regime, higher-order structure in the MDF cannot be robustly constrained, and a Gaussian has therefore been widely adopted as a descriptive, rather than physical, model in the UFD literature. Producing a truly Gaussian distribution with a dispersion approaching $\sim0.7$ dex is challenging within standard chemical evolution frameworks. Rather, a statistically preferred Gaussian likely indicates either unresolved internal structure that cannot yet be recovered with available data or an extreme degree of metal mixing \citep{frebel_chemical_2012, webster_segue_2016}. Consistent with this interpretation, chemical-evolution models (leaky box, pre-enriched, extra gas) offer no statistical improvement over a single-Gaussian MDF, highlighting both the limitations of current models in this regime and the need for larger, higher-precision samples to resolve the true form of the MDF.

Turning to the one-shot enrichment scenario, our results are compatible with, but not uniquely indicative of, this framework. Prior high-resolution work found persistently high [$\alpha$/Fe] ratios in Segue~1 stars with no decline at higher [Fe/H], implying chemical contributions solely from core-collapse supernovae, with no detectable Type Ia enrichment \citep{frebel_segue_2014}. This picture aligns with Segue~1 retaining signatures of its earliest Population III and massive-star supernovae. While the MDF alone does not directly encode the star formation history, its shape can be interpreted through chemical-evolution arguments. In particular, \citet{webster_segue_2016} showed that in a one-shot enrichment scenario, a broad MDF can result from inhomogeneous enrichment by a small number of early supernovae rather than prolonged star formation. The lack of distinct substructure, together with uniformly high [$\alpha$/Fe] ratios, is consistent with a brief early enrichment episode followed by rapid quenching, though the MDF alone cannot uniquely distinguish among scenarios.

For the multiple-burst scenario, our data provide no evidence for a bimodal or trimodal MDF: a single Gaussian is preferred over both multi-component models (Table~\ref{tab:mdf_fits}, Figure~\ref{fig:mdf_models}). While the 7-star sample in \citet{frebel_segue_2014} suggested three distinct metallicity peaks, our larger sample of $N=45$ stars does not reproduce such substructure. To quantify the sensitivity of our data to such substructure, we simulated whether a truly trimodal MDF with peaks at [Fe/H] $\approx -1.8, -2.3$, and $-3.3$ (as suggested in the 7-star MDF in \citealt{frebel_segue_2014}) could be recovered given observational uncertainties of 0.3 dex. We generated mock samples of increasing size from a trimodal Gaussian distribution, added average metallicity uncertainties, and fit single- and three-component Gaussian mixture models to compare their AIC. The results show that with a sample comparable to ours (N = 45), the trimodal distribution is only correctly preferred $\sim44\%$ of the time, and $>90\%$ reliability is only reached with $\geq 150-200$ stars. To reach a level where the trimodal distribution is correctly preferred in $\geq 99.9\%$ cases requires $\geq250$ stars, which may be observationally prohibitive for Segue~1 given its 250th brightest star falls at $g \sim 26.5$. 

The updated MDF of Segue~1, considered alongside detailed chemical abundance measurements, points to brief early star formation with limited chemical evolution. This timing is independently supported by resolved-star photometry: from a uniform reanalysis of archival HST imaging, \citet{durbin_hst_2025} find that Segue~1 formed 50\% and 80\% of its stellar mass by $\tau_{50} = 13.18^{+0.04}_{-1.56}$ and $\tau_{80} = 13.08^{+0.01}_{-4.36}$~Gyr ago, consistent with star formation ending before reionization. However, the large uncertainties indicate that current photometry cannot resolve the short timescales that separate the three scenarios considered here. This picture is also broadly consistent with the one-shot enrichment scenario, in which Segue~1’s stellar population preserves the signatures of the earliest supernovae and traces the earliest phase of galaxy formation \citep{frebel_chemical_2012, webster_star_2015, webster_segue_2016}. However, the smooth, Gaussian-like MDF alone does not definitively rule out the chemical enrichment pathways often seen in more massive UFDs, and our small sample size limits our ability to distinguish between models. 

Similar studies in other UFDs (e.g., Boötes I; \citealt{lai_feh_2011}) show that multiple analytic chemical evolution models can remain statistically viable even with comparable sample sizes. Thus, while the one-shot scenario remains a compelling framework, additional spectroscopic data are needed to distinguish whether Segue~1 experienced a strictly single burst, clustered star formation, or a more extended but still short-lived episode of star formation ending before reionization ($z\sim$ 6,~$\lesssim1$~Gyr; \citealt{brown_quenching_2014, durbin_hst_2025}); shorter-timescale bursty substructure would likely remain unresolved given current MDF uncertainties, even if metal mixing were inefficient. 

\subsection{Constraints on Population III Survivors in Segue~1}

UFDs are among the best places to search for surviving Population III (Pop~III) stars. Their extremely low metallicities and simple star formation histories make them natural laboratories to preserve the signatures of the first stars, whereas more massive galaxies have undergone repeated enrichment events that would obscure or dilute any primordial survivors \citep{hartwig_constraining_2015, magg_predicting_2018, rossi_ultra-faint_2021, rossi_hidden_2025}. The most metal-poor star in our sample, SDSSJ100643.47+160604.5, has ${\rm [Fe/H]} = -3.94 \pm 0.37$, placing it in the regime where abundance patterns preserving signatures of Pop~III supernovae are plausible \citep{karlsson_pregalactic_2013, chiaki_seeding_2019}. The large uncertainty inherent to Ca~\textsc{II}~K metallicities at low [Fe/H] means that high-resolution spectroscopic follow-up would be needed to confirm its metallicity and determine whether its abundance pattern bears the signatures of Pop~III enrichment. Such follow-up has proven valuable in other UFDs: \citet{chiti_enrichment_2026} recently reported detailed abundances of a star with ${\rm [Fe/H]} < -4.63$ in a UFD whose chemical abundance pattern strongly suggests preservation of Pop~III enrichment signatures, demonstrating that stars in this metallicity regime in UFDs can serve as direct fossil records of the first stars.

While all of the stars in our sample contain metals, we use the updated MDF to place new constraints on the presence of long-lived Pop~III stars in Segue~1. We derive upper limits on the number of Pop~III stars in Segue~1 using a one-sided Clopper--Pearson confidence interval for a binomial distribution, with $k=0$ detections of stars below [Fe/H]~$=-5$ in a sample of $N=45$ members \citep{clopper_use_1934}. We adopt ${\rm [Fe/H]} = -5$ as the Pop~III threshold, corresponding to the critical metallicity for dust cooling, and because metallicities below this level are difficult to constrain at the resolution of our data \citep{schneider_first_2012}. The fraction of surviving Pop~III stars has been estimated by \citet{magg_predicting_2018}, who model low-mass Pop~III star formation using merger trees from N-body simulations, assuming a logarithmically flat initial mass function (IMF) in the range $0.6-150 M_{\odot}$ and metal enrichment from neighboring supernovae to regulate the transition from Pop~III to Pop~II star formation. Under these assumptions, they place an upper limit on the fraction of Pop~III survivors in Segue~1 at 1.5\%. 

Our approach provides a bound on the surviving Pop~III star fraction $f_{\rm Pop~III}$, such that the probability of obtaining zero detections is less than 5\% (for $95\%$ confidence) or 0.27\% (for $99.7\%$ confidence) if the true fraction were larger than the quoted limit. Given our sample of $N=45$ confirmed Segue~1 member stars with no detections below [Fe/H]~$=-5$, we place an upper limit on the fraction of Pop~III stars of $f_{\rm Pop~III} \leq 0.064$ ($2\sigma$ confidence).  At the $3\sigma$ level, we constrain $f_{\rm Pop~III} \leq 0.109$. 

We also calculate the required number of stars needed for a prospective detection through a similar estimate from the binomial probability $P(\geq 1~{\rm detection}) = 1-(1-f_{\rm Pop~III})^N$. If the true Pop~III star fraction was 1\% (e.g., \citealt{magg_predicting_2018, rossi_ultra-faint_2021}), one would need to observe at least $N \gtrsim 299$ member stars to have a 95\% chance of detecting at least one such star, or $N \gtrsim 589$ for a 99.7\% confidence level. Given its total stellar mass of $M_{*} \approx 500 M_\odot$ and a standard \citet{kroupa_variation_2001} IMF, Segue~1 is estimated to contain only a few thousand stars; however, the 300th brightest star falls at $g \sim 27$, well beyond the reach of current multi-object spectrographs.

This exercise highlights that, even under optimistic assumptions, constraining the form of the Pop~III IMF in an ultra-faint system like Segue~1 would require stellar spectroscopy for several hundred stars. The fraction of surviving Pop~III stars is directly tied to the low-mass end of the Pop~III IMF: a higher fraction of survivors implies a less top-heavy Pop~III IMF with more low-mass ($\lesssim 0.8\,M_\odot$) Pop~III stars, while zero detections are consistent with a predominantly massive, top-heavy Pop~III IMF in which few or no low-mass stars formed. Placing meaningful constraints on this low-mass end of the Pop~III IMF through direct detection of Pop~III survivors would therefore require stellar spectroscopy for several hundred member stars in a system like Segue~1. 

\section{Conclusions} \label{sec:conclusion}

Segue~1 remains one of the most primitive galaxies known, offering a rare glimpse into the earliest phases of galaxy formation. This work presents an updated MDF for Segue~1, with a mean metallicity of $\langle\mathrm{[Fe/H]}\rangle = -2.52 \pm 0.10$ and an intrinsic dispersion of $\sigma_{\mathrm{[Fe/H]}} = 0.59 \pm 0.06$ dex. This distribution does not reproduce the trimodal structure reported by \citet{frebel_segue_2014}. With a six-fold larger sample size compared to previously published metallicities, this observed single-peaked MDF supports the idea that Segue~1 may not have formed as three distinct star clusters, but instead from a brief, continuous star formation process. The measured metallicity spread further reinforces Segue~1’s classification as a galaxy, not a star cluster, as it demonstrates retention of supernova ejecta.

Chemical evolution models (leaky box, pre-enriched, and extra gas) do not provide a better fit to the MDF than a single Gaussian. Additionally, the absence of stars below [Fe/H] $= -5$ enables us to set an upper limit of $f_{\mathrm{Pop\,III}} \lesssim 0.064$ at $2\sigma$ confidence on the surviving Population~III fraction, underscoring the difficulty of constraining the Pop~III IMF in systems with such small stellar populations.

Future observations will be key to refining these constraints. Extremely Large Telescopes (ELTs) will enable spectroscopy of larger samples of faint main-sequence stars in Segue~1, allowing for improved statistics of its MDF. ELTs will gain $\sim$2 magnitudes in depth over current facilities (to $g\sim 24$), roughly doubling the number of accessible Segue~1 members to $\sim$100 stars — still a factor of $\sim$2.5 short of the $\geq$250 spectra needed to distinguish a trimodal MDF. Given the limitations of current instruments, this is the most comprehensive spectroscopic MDF available for Segue~1 to date. However, improved metallicity calibration could expand the range of accessible stars and reduce metallicity uncertainties. 

The case of Segue~1 also illustrates the large observational effort required to extract even limited information about the faintest galaxies. In the era of the Vera C. Rubin Observatory, the discovery space for ultra-faint systems will expand dramatically. While wide-field surveys will revolutionize the census of the faintest galaxies, obtaining the detailed stellar abundances needed to understand their formation and evolution will demand a new level of spectroscopic capability and investment.

Overall, this work demonstrates the importance of expanding spectroscopic samples in the faintest galaxies: while Segue~1’s MDF appears simple, it has implications for understanding the earliest stages of galaxy formation, chemical evolution, and the role of the smallest dark matter halos in the early universe.

\section{Acknowledgments}
D.B. and A.P.J. acknowledge support from the National Science Foundation under grant AST-2307599. This paper made use of NASA’s Astrophysics Data System
Bibliographic Services. This work was completed in part with resources provided by the University of Chicago’s Research Computing Center.

Some of the data presented herein were obtained at Keck Observatory, which is a private 501(c)3 non-profit organization operated as a scientific partnership among the California Institute of Technology, the University of California, and the National Aeronautics and Space Administration. The Observatory was made possible by the generous financial support of the W. M. Keck Foundation.

The authors wish to recognize and acknowledge the very significant cultural role and reverence that the summit of Maunakea has always had within the Native Hawaiian community. We are most fortunate to have the opportunity to conduct observations from this mountain.

This research has made use of the Keck Observatory Archive (KOA), which is operated by the W. M. Keck Observatory and the NASA Exoplanet Science Institute (NExScI), under contract with the National Aeronautics and Space Administration.

This research used the facilities of the Canadian Astronomy Data Centre operated by the National Research Council of Canada with the support of the Canadian Space Agency and uses observations obtained at the Canada-France-Hawai\okina i Telescope (CFHT) which is operated by the National Research Council of Canada, the Institut National des Sciences de l'Univers of the Centre National de la Recherche Scientifique of France, and the University of Hawai\okina i. CFHT is located on Maunakea on Hawai\okina i Island, a mountain of considerable cultural, natural, and ecological significance. Maunakea is a sacred site to Native Hawaiians, also known as K\={a}naka \okina\={O}iwi. Quality observations are made possible by relentless effort of the entire staff at Canada-France-Hawai\okina i Telescope. Based on observations obtained with MegaPrime/MegaCam, a joint project of CFHT and CEA/DAPNIA.

Funding for the SDSS and SDSS-II has been provided by the Alfred P. Sloan Foundation, the Participating Institutions, the National Science Foundation, the U.S. Department of Energy, the National Aeronautics and Space Administration, the Japanese Monbukagakusho, the Max Planck Society, and the Higher Education Funding Council for England. The SDSS Web Site is http://www.sdss.org/.

The SDSS is managed by the Astrophysical Research Consortium for the Participating Institutions. The Participating Institutions are the American Museum of Natural History, Astrophysical Institute Potsdam, University of Basel, University of Cambridge, Case Western Reserve University, University of Chicago, Drexel University, Fermilab, the Institute for Advanced Study, the Japan Participation Group, Johns Hopkins University, the Joint Institute for Nuclear Astrophysics, the Kavli Institute for Particle Astrophysics and Cosmology, the Korean Scientist Group, the Chinese Academy of Sciences (LAMOST), Los Alamos National Laboratory, the Max-Planck-Institute for Astronomy (MPIA), the Max-Planck-Institute for Astrophysics (MPA), New Mexico State University, Ohio State University, University of Pittsburgh, University of Portsmouth, Princeton University, the United States Naval Observatory, and the University of Washington.

\vspace{3pt} 

\facilities{Keck:I (LRIS), CFHT (MegaCam), \textit{Sloan}.}

\software{\texttt{numpy} \citep{harris2020array}, \texttt{matplotlib} \citep{Hunter:2007}, \texttt{scipy}  \citep{2020SciPy-NMeth}, \texttt{pandas} \citep{the_pandas_development_team_2025_17229934}, \texttt{astropy} \citep{astropy:2013, astropy:2018, astropy:2022}.}

\clearpage
\bibliography{references}{}
\bibliographystyle{aasjournal}

\end{document}